\documentclass[]{spie}  %>>> use for US letter paper
\usepackage{amsmath,amsfonts,amssymb}
\usepackage{graphicx}
\usepackage[colorlinks=true, allcolors=blue]{hyperref}
\usepackage{float}

\title{Focus diverse phase retrieval test results on broadband continuous wavefront sensing in space telescope applications}

\author[a]{Hyukmo Kang}
\author[b]{Kyle Van Gorkom}
\author[a]{Meghdoot Biswas}
\author[a,b]{Daewook Kim}
\author[b]{Ewan S. Douglas}

\affil[a]{James C. Wyant College of Optical Sciences, University of Arizona, 1630 E. University Blvd., Tucson, AZ 85721 USA}
\affil[b]{Department of Astronomy, University of Arizona, 933 N. Cherry Ave., Tucson, AZ, 85721 USA}

\authorinfo{Further author information: (Send correspondence to Hyukmo Kang)\\Hyukmo Kang: E-mail: hkang04@arizona.edu\\  Daewook Kim: E-mail: dkim@optics.arizona.edu \\ Ewan S.Douglas: E-mail: douglase@arizona.edu}

\begin{document} 
\maketitle

\begin{abstract}
Continuous wavefront sensing benefits space observatories in on-orbit optical performance maintenance. To measure the phase of a wavefront, phase retrieval is an attractive technique as it uses multiple point spread function (PSF) images that are acquired by the telescope itself without extra metrology systems nor complicated calibration. The focus diverse phase retrieval utilizes PSFs from predetermined defocused positions to enhance the dynamic range of the algorithm. We describe an updated visible light active optics testbed with the addition of a linear motorized focus stage. The performance of the phase retrieval algorithm in broadband is tested under various cases. While broadband pass filters have advantages in higher signal-to-noise ratio (SNR), the performance of phase retrieval can be restricted due to blurred image caused by diffraction and increased computing cost. We used multiple bandpass filters (10~nm, 88~nm, and 150~nm) and investigated effects of bandwidth on the accuracy and required image acquisition conditions such as SNR, reaching accuracies below 20~nm RMS wavefront error at the widest bandwidth. We also investigated the dynamic range of the phase retrieval algorithm depending on the bandwidth and required amount of defocus to expand dynamic range. Finally, we simulated the continuous wavefront sensing and correction loop with a range of statistically generated representative telescope disturbance time series to test for edge cases.  
\end{abstract}

\keywords{wavefront sensing, phase retrieval, space telescope, broadband}

\section{INTRODUCTION}
\label{sec:intro}  
Phase retrieval is a method for recovering phase information in optical systems using one or more intensity measurements along with prior information to limit the solution space. As phase retrieval employs point-spread function (PSF) images at the focal plane, it does not require additional metrology systems which is beneficial for space telescopes due to reduced hardware complexity and weight. Consequently, phase retrieval is extensively utilized in missions from the Hubble Space Telescope to the James Webb Space Telescope to assess optical performance\cite{fienup1993hubble, dean2006phase}.

Focus diversity phase retrieval (FDPR) employs multiple defocused PSFs to resolve ambiguities in the pupil-plane phase solution. The usage of multiple PSFs enhances robustness against noise and improving sensitivity to specific spatial frequencies\cite{fienup2013phase,dean2003diversity,jurling2014applications}. 

We continued the development of the phase retrieval testbed for continuous wavefront sensing and control\cite{kang2023}. As part of this progress, we adopted a halogen lamp as the light source to test how the algorithm's performance varies with different bandpass filters. In this report, we present current testbed results that evaluate the algorithm's performance across various bandpass filters. We analyzed how the success ratio changes with bandpass, defocus range, and signal-to-noise ratio. Finally, we simulated telescope drift scenarios to assess whether we can maintain the root-mean-square (RMS) wavefront error below 20~nm.

\section{Testbed setup}

\begin{figure}[H]
\centering\includegraphics[width=10cm]{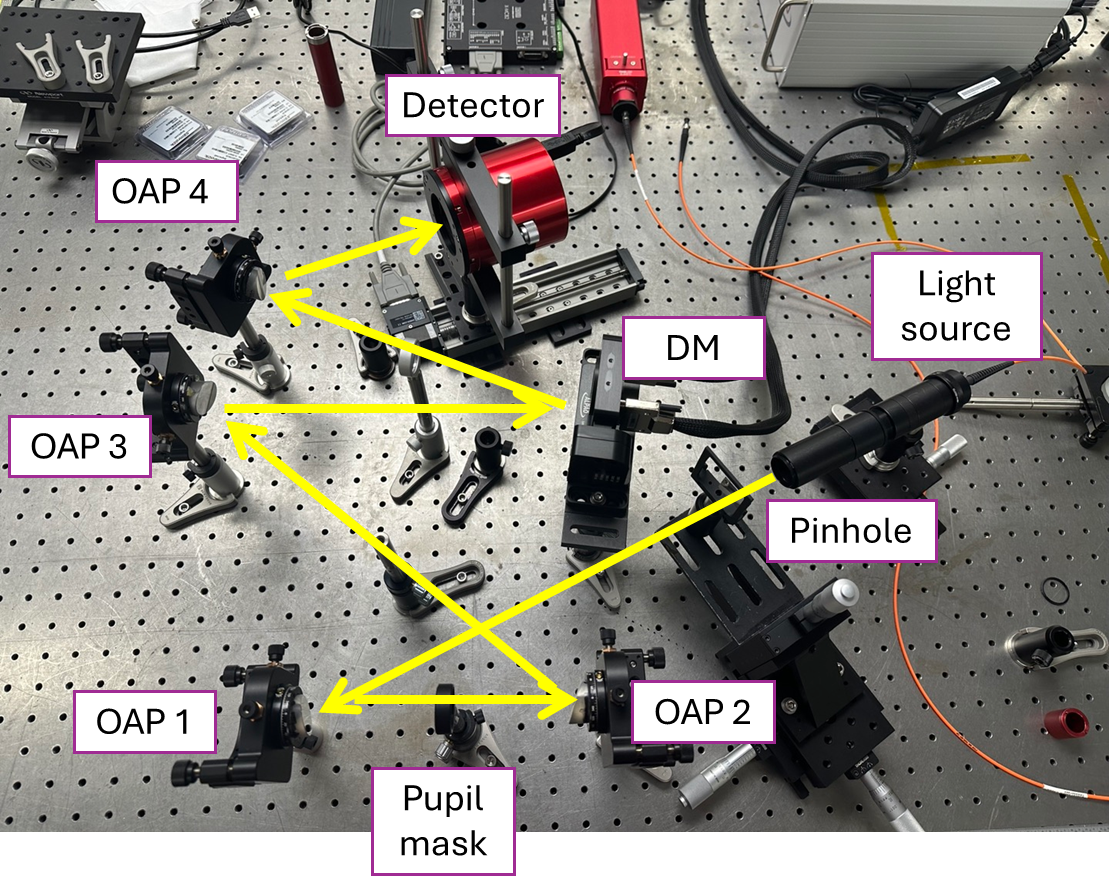}
\caption{Layout of the testbed. Yellow arrows represent the path of the light. }
\label{fig:testbed_layout}
\end{figure}

Current testbed setup is shown at Fig.~\ref{fig:testbed_layout}. We updated the setup from our previous testbed for broadband phase retrieval tests \cite{kang2023}. Firstly, we replaced the light source fron HeNe laser to Halogen lamp with bandpass filters. The bandpass filters we used have central wavelengths / bandwidths of 633~nm/10~nm, 550~nm/88~nm, and 650~nm/150~nm. The beam from the fiber tip is imaged and pass through the pinhole. We placed the detector on a motorized linear stage (Zaber X-LHM150A) to create defocused PSFs.

The FDPR code was re-written to enable phase retrieval on broadband images. The bandpass of interest is discretized into individual wavelengths, and the electric field for each wavelength is independently propagated to the defocused planes via Fresnel propagation implemented by matrix Fourier transforms (MFT) and incoherently summed to produced the modeled intensity images. The pupil-plane optical path difference (OPD) is parameterized by either a modal basis set (e.g., Zernike polynomials) or individual pixels. The forward model is written in \texttt{JAX}\cite{jax2018github}, which allows automatic differentiation of the forward propagation model. The OPD map is estimated by use of optimization routines in the \texttt{jaxopt} codebase\cite{jaxopt_implicit_diff}.

\section{Results}
To check the performance changes of the FDPR algorithm according to bandpass, we generated 101 random surface errors on the deformable mirror (DM) with peak-to-valley (PV) wavefront error (WFE) ranging from 1 to 5.5~$\lambda$ using Zernike coefficients from defocus (Z4) to spherical aberration (Z11), and measure the DM surface using interferometer and FDPR with different bandpass filters.

\begin{figure}[H]
\centering\includegraphics[width=15cm]{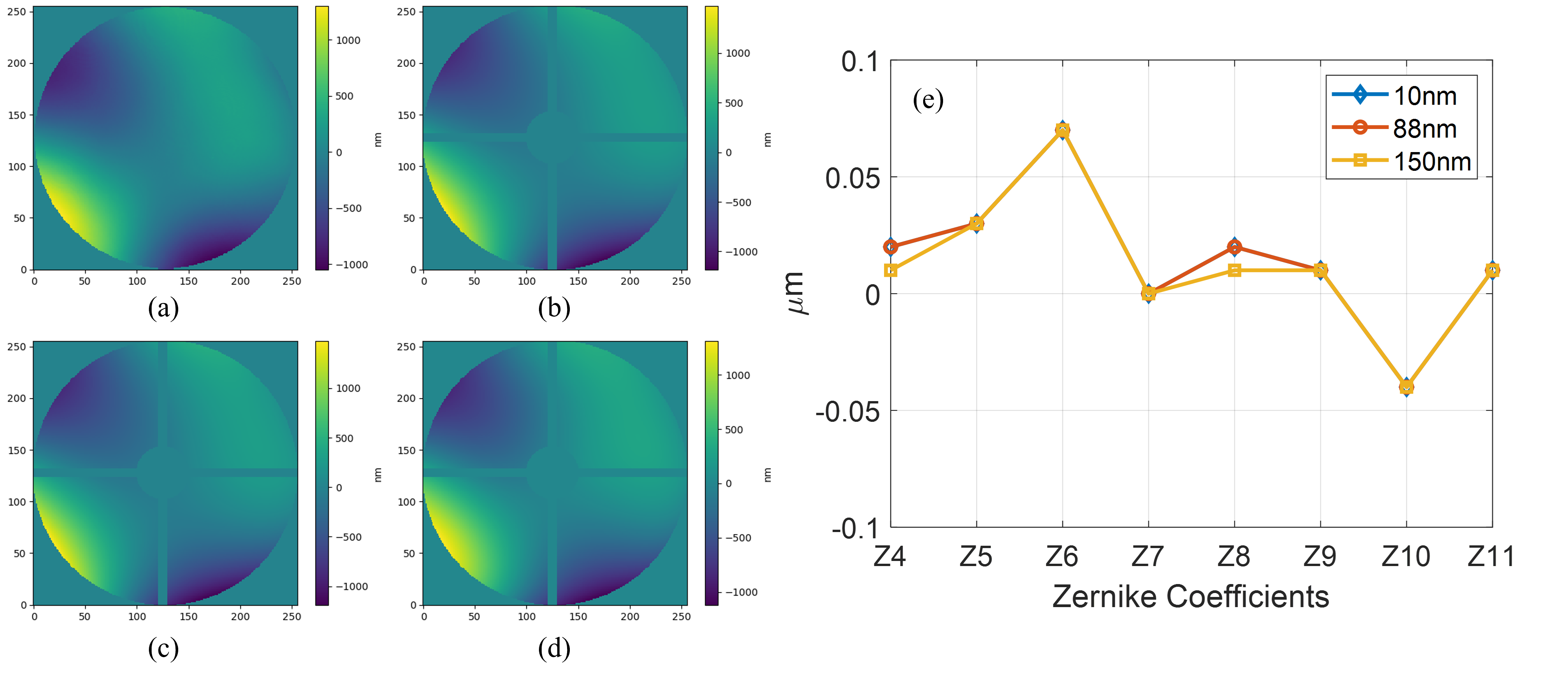}
\caption{Measured shape of deformable mirror from (a) Interferometer (b) phase retrieval with 10~nm (c) 88~nm and (d) 150~nm of bandpass filters. (e) Difference between the Zernike coefficients of the interferometer and PR. The results do not change significantly even when the bandpass varies. 
}
\label{fig:pr_result_bandpass}
\end{figure}

Fig.~\ref{fig:pr_result_bandpass} shows the measurement results for one case. The PV WFE measured by the interferometer is approximately 2353 nm (about 3.7~$\lambda$, Fig.~\ref{fig:pr_result_bandpass}(a)). The results of the FDPR using various bandpass filters are 2660~nm (10~nm, Fig.~\ref{fig:pr_result_bandpass}(b)), 2671~nm (88~nm, Fig.~\ref{fig:pr_result_bandpass}(c)), and 2434~nm (150~nm, Fig.~\ref{fig:pr_result_bandpass}(d)), which are not significantly different from the interferometer results (with a maximum difference of 318~nm, or about 0.5~$\lambda$). Additionally, the differences in the measured Zernike coefficients (Fig.~\ref{fig:pr_result_bandpass}(e)) also showed similar results across the different bandpasses, indicating that the variations in bandpass do not have a significant impact on the performance of the PR algorithm.

\begin{figure}[H]
\centering\includegraphics[width=17cm]{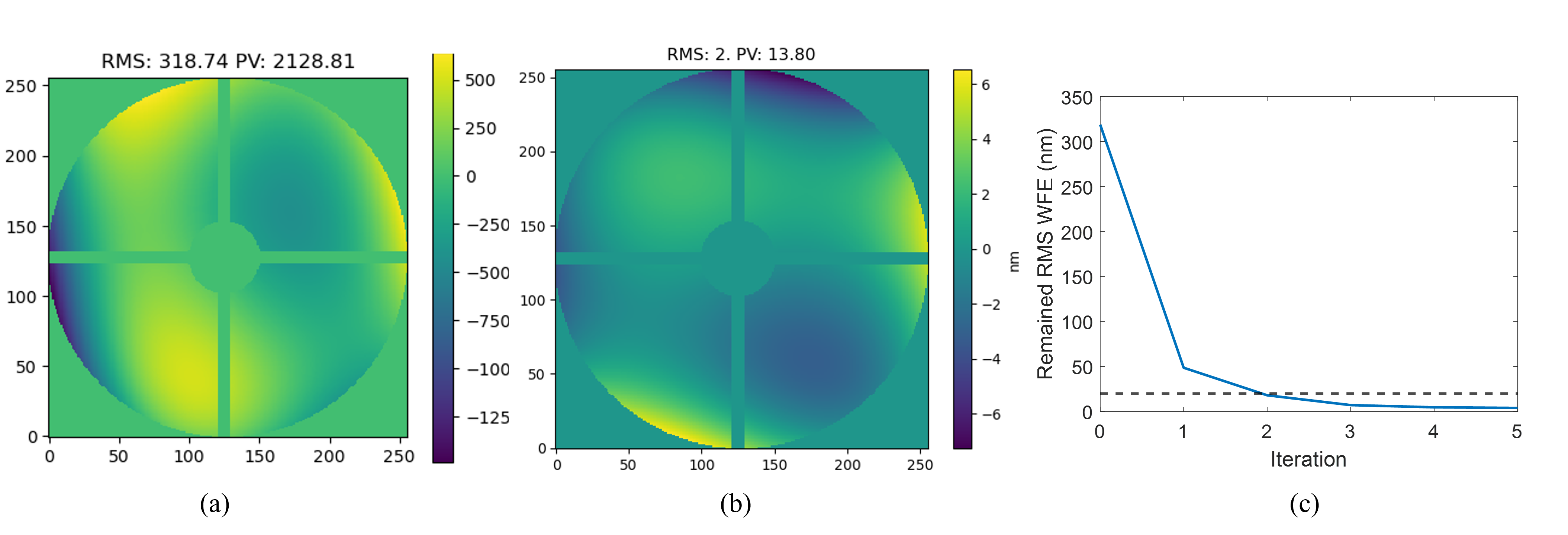}
\caption{(a) Initial surface. (b) After 5 times of iterations of correction through phase retrieval. Note differing colorbar scales. (c) Changes in the remaining RMS WFE with iterations. Residual RMS WFE is under 20~nm after 2 iterations.}
\label{fig:one_closed_loop}
\end{figure}

In this experiment, we assumed that the PR results were successful in measuring surface error when the difference from the interferometer results was less than 1 wave. This is because, in this experiment, we did not account for additional measurement errors that could arise from moving the detector to create defocus. As shown in Fig.~\ref{fig:one_closed_loop}, even when considering measurement errors, if the residual WFE is small after the initial measurement and correction, applying iterative PR can almost eliminate the shape error.

\begin{figure}[H]
\centering\includegraphics[width=11cm]{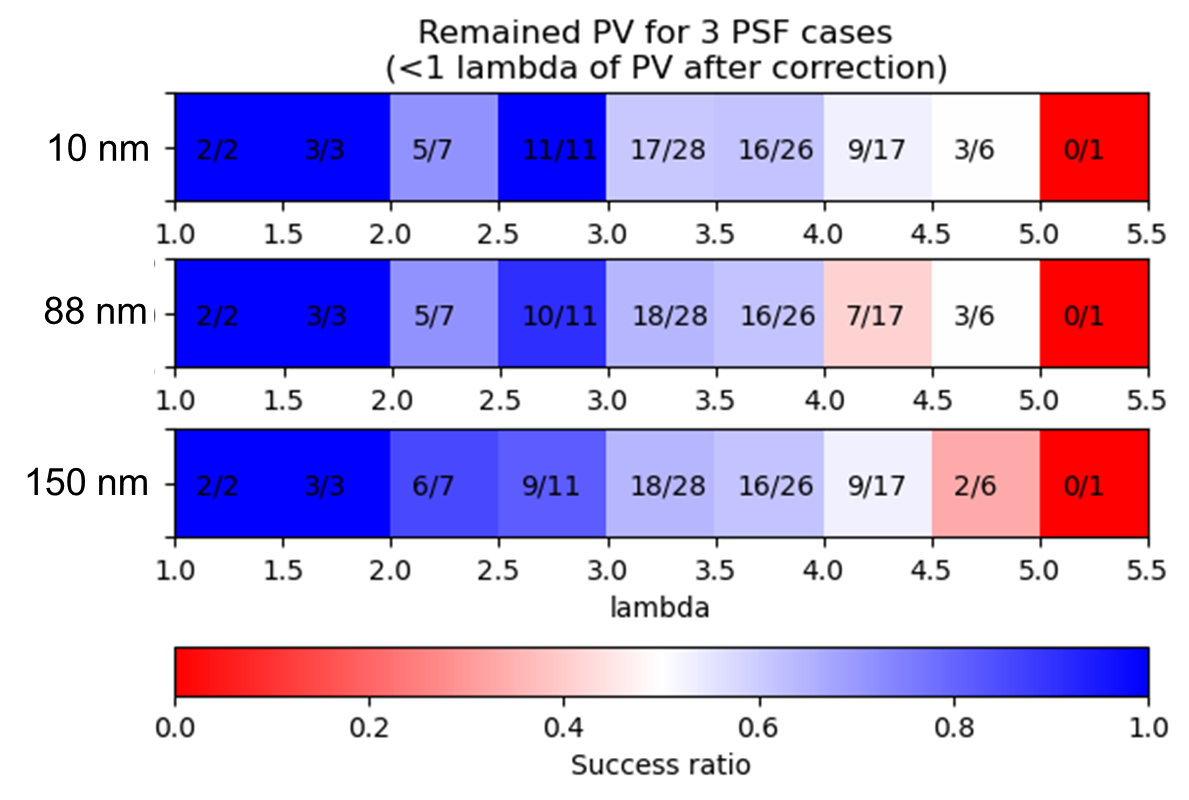}
\caption{Success ratio by bandpass. Even with an increase in bandpass, there is no significant difference in the results of phase retrieval.
}
\label{fig:pr_result_stats_bandpass}
\end{figure}

Fig.~\ref{fig:pr_result_stats_bandpass} shows the success ratio of each bandpass filter case. Similar to the results above, we observe that the results do not vary significantly across the different bandpasses for most PV ranges. This indicates that the bandpass does not have a substantial impact on the performance of our current FDPR algorithm. Furthermore, using a wider bandpass filter, such as the 150~nm filter, may be advantageous for practical FDPR applications, as it allows for the collection of more light. Since we found no significant issues with using the 150~nm filter, we decided to conduct all future experiments using the 150~nm filter.

\begin{figure}[H]
\centering\includegraphics[width=15cm]{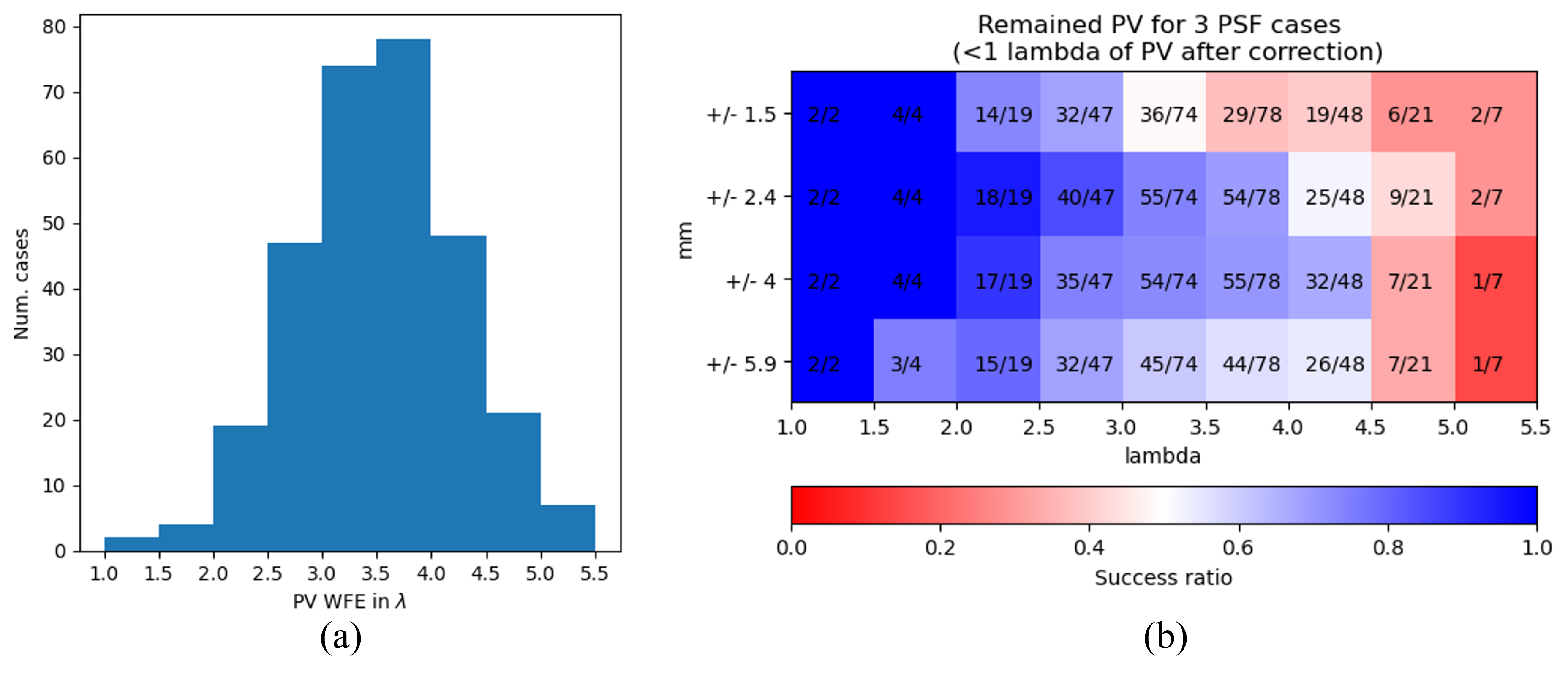}
\caption{(a) Distribution of PV WFE that we used for the experiment. (b) Phase retrieval success rate depending on the amount of defocus. When the defocus is ±2.4mm or ±4mm, it shows the best results in our main area of interest.}
\label{fig:defocus_result}
\end{figure}

Next, we aimed to investigate the performance changes of FDPR according to the defocus range. Similar to the previous experiment, we generated 300 random surface errors with PV WFE ranging from 1 to 5.5 lambda using Zernike coefficients. Fig.~\ref{fig:defocus_result} shows the success ratios when using defocus values of 0.075, 0.12, 0.2, and 0.3 um of RMS, respectively. As a result, we found that for PV WFE up to 4~$\lambda$, FDPR performed best at 0.12 (±2.4~mm of detector shift) and 0.2 (±4~mm of detector shift) $\mu$m of RMS in defocus.

\begin{figure}[H]
\centering\includegraphics[width=15cm]{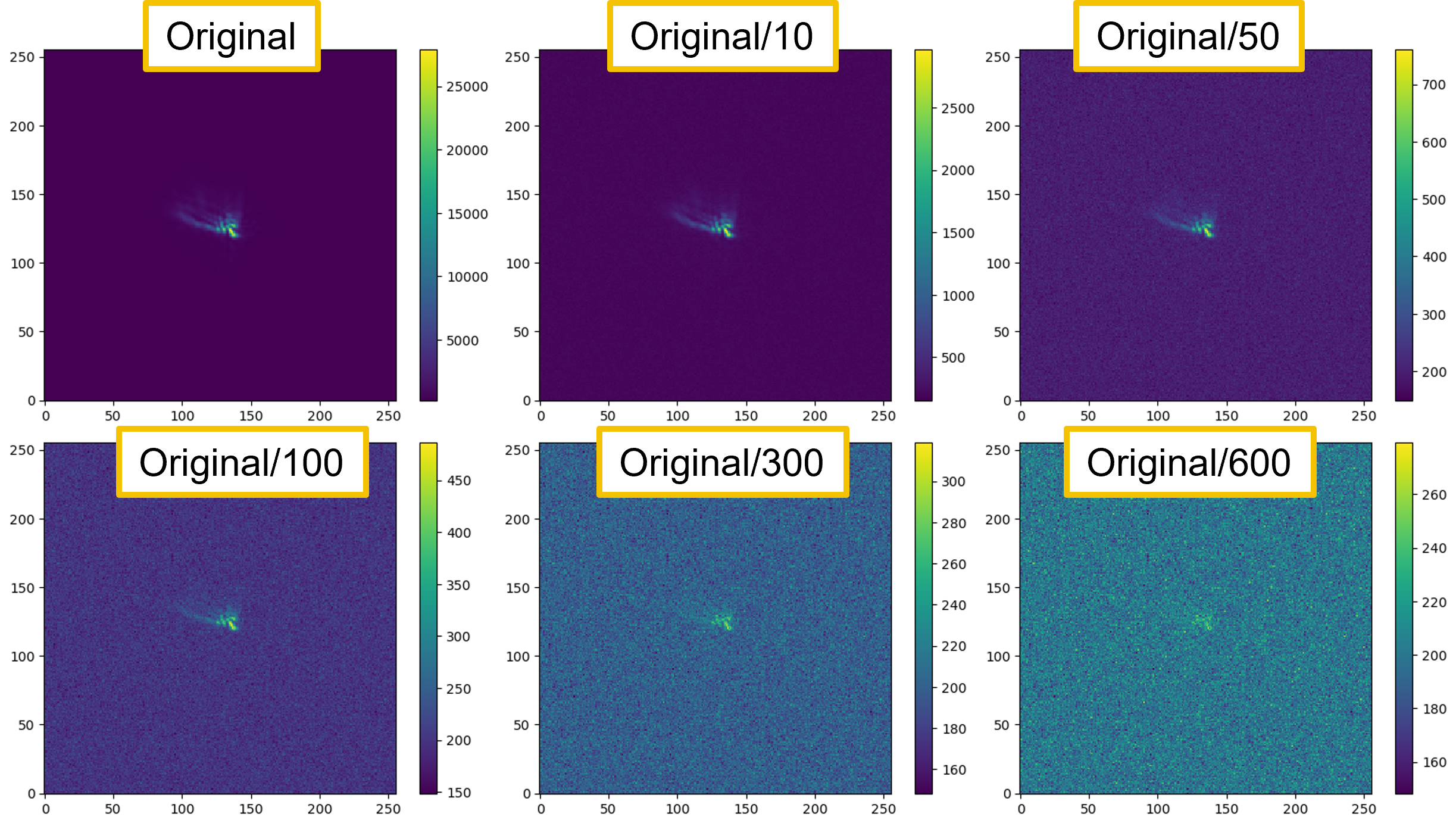}
\caption{Example image of noise-added PSF images. The original PSF is from 3.5~$\lambda$ of PV WFE case.}
\label{fig:noise_example}
\end{figure}

In the following step, we aimed to examine the performance changes of FDPR according to the signal-to-noise ratio (SNR). We decided to use the case with a bandpass of 150~nm and a defocus range of ±2.4~mm (0.12~$\mu$m), which showed good results in the previous experiments. To create a noisy PSF using the same measurement data, we divided the measured PSF image by a constant value and then added the same shot noise (see Fig.~\ref{fig:noise_example}). Here, since we assumed the same amount of shot noise, we can expect that the larger the intensity of the PSF (i.e., the smaller the assigned PV WFE), the lesser the impact of the shot noise.

\begin{figure}[H]
\centering\includegraphics[width=15cm]{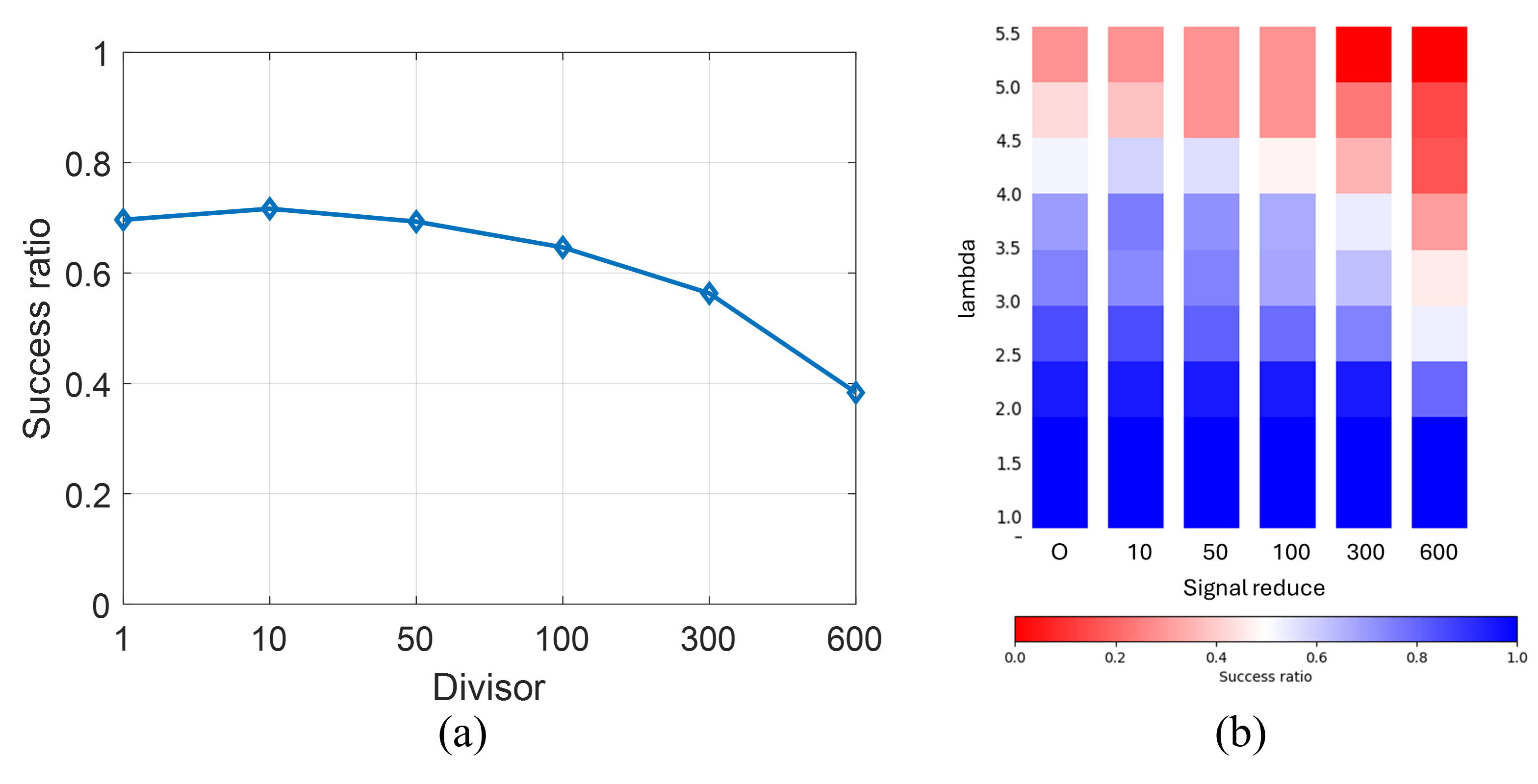}
\caption{(a) Overall success ratio by divisor. (b) success ratio by the input PV WFE. The trend of graphs show that the dynamic range of FDPR can be limited by the SNR of PSF images.}
\label{fig:snr_result}
\end{figure}

Fig.~\ref{fig:snr_result}(a) shows the success rate of FDPR by divisor, while Fig.~\ref{fig:snr_result}(b) presents the success rate for each divisor given the assigned PV WFE. As expected, when the input PV WFE is small, there is little impact from noise. However, in cases with high PV WFE, we can see that the success rate significantly decreases as the signal contrast diminishes (i.e., as the divisor increases). We expect that these simulation and measurement results will provide valuable reference material for calculating the required exposure time for telescopes.

\begin{figure}[H]
\centering\includegraphics[width=15cm]{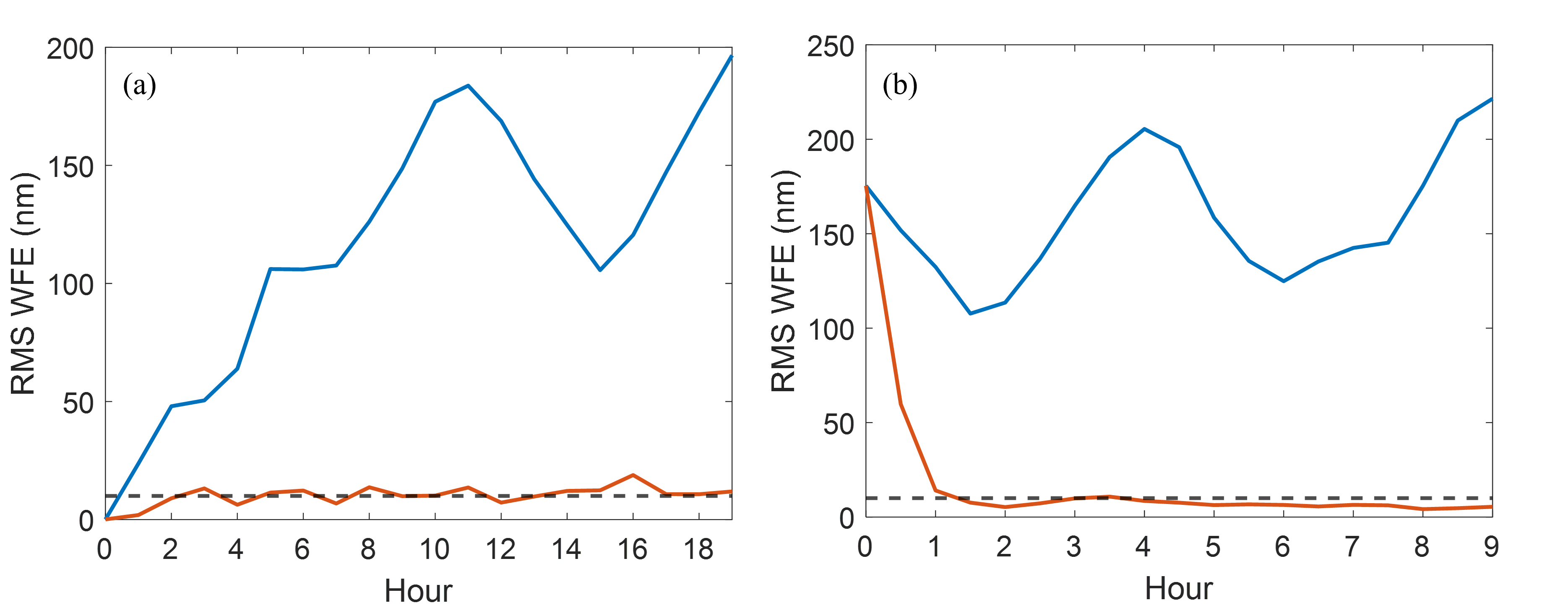}
\caption{Residual RMS WFE for different temporal lag cases. The synthetic data is from the telescope drifting case study. The blue line represents the case without wavefront correction, while the red line shows the cases of (a) one hour and (b) 30-minute temporal gap for wavefront correction. The remained WFE is kept below 20~nm in the case of the 30-minute interval.
}
\label{fig:closed_loop}
\end{figure}

Finally, we confirmed whether we could maintain the desired performance using a broadband phase retrieval. The data used for the experiment was based on the synthetic data in a telescope drifting scenario\cite{douglas2023}. Fig.~\ref{fig:closed_loop}(a) assumes WFE measurements are taken hourly, while (b) assumes measurements are taken every 30 minutes. In both cases, the residual WFE was maintained at approximately 20~nm. However, as expected, the case with less frequent measurements (Fig.~\ref{fig:closed_loop}(a)) showed a tendency for the remained RMS WFE to be higher, despite a lower initial RMS WFE compared to the more frequent measurement case. Similar to the SNR experimental results above, we anticipate that this experimental data can inform the determination of the wavefront sensing and correction cycle for future telescopes.

\section{Conclusions and Future work}
In this report, we examined the progress of the FDPR testbed for continuous wavefront sensing and how the performance of FDPR may vary in the case of broadband. The experimental results confirmed that there was no significant difference in accuracy between the 150~nm bandwidth and the 10~nm case. Additionally, it was found that using defocused PSF in the range of 0.12 to 0.2~$\mu$m is efficient for correcting WFE at the PV level of 4~$\lambda$. Moreover, the experimental results and simulations related to SNR and closed-loop conditions can assist in calculating the exposure time and determining the correction cycle based on the expected WFE during telescope operations.

In this experiment, WFE was generated using only a deformable mirror. The goal of the next experiment is to create a telescope simulator that not only implements shape errors with a deformable mirror but also incorporates rigid body motion to simulate misalignment of the actual telescope, allowing us to assess the performance of FDPR under these conditions.

\acknowledgments % equivalent to \section*{ACKNOWLEDGMENTS}       
 
Portions of this research were supported by funding from the Technology Research Initiative Fund (TRIF) of the Arizona Board of Regents and by generous anonymous philanthropic donations to the Steward Observatory of the College of Science at the University of Arizona. 

% References
\bibliography{report} % bibliography data in report.bib

\begin{thebibliography}{1}

\bibitem{fienup1993hubble}
Fienup, J.~R., Marron, J.~C., Schulz, T.~J., and Seldin, J.~H., ``Hubble space telescope characterized by using phase-retrieval algorithms,'' {\em Applied optics}~{\bf 32}(10),  1747--1767 (1993).

\bibitem{dean2006phase}
Dean, B.~H., Aronstein, D.~L., Smith, J.~S., Shiri, R., and Acton, D.~S., ``Phase retrieval algorithm for jwst flight and testbed telescope,'' in [{\em Space telescopes and instrumentation I: optical, infrared, and millimeter}{\nolinebreak\hspace{0.1em}]},   {\bf 6265},  314--330, SPIE (2006).

\bibitem{fienup2013phase}
Fienup, J.~R., ``Phase retrieval algorithms: a personal tour,'' {\em Applied optics}~{\bf 52}(1),  45--56 (2013).

\bibitem{dean2003diversity}
Dean, B.~H. and Bowers, C.~W., ``Diversity selection for phase-diverse phase retrieval,'' {\em JOSA A}~{\bf 20}(8),  1490--1504 (2003).

\bibitem{jurling2014applications}
Jurling, A.~S. and Fienup, J.~R., ``Applications of algorithmic differentiation to phase retrieval algorithms,'' {\em JOSA A}~{\bf 31}(7),  1348--1359 (2014).

\bibitem{kang2023}
Kang, H., Gorkom, K.~V., Johnson, J., Singelstad, O., Goldtooth, A., Kim, D., and Douglas, E.~S., ``{Focus diverse phase retrieval testbed development of continuous wavefront sensing for space telescope applications},'' in [{\em Astronomical Optics: Design, Manufacture, and Test of Space and Ground Systems IV}{\nolinebreak\hspace{0.1em}]},  Hull, T.~B., Kim, D., and Hallibert, P., eds.,  {\bf 12677},  126770G, International Society for Optics and Photonics, SPIE (2023).

\bibitem{jax2018github}
Bradbury, J., Frostig, R., Hawkins, P., Johnson, M.~J., Leary, C., Maclaurin, D., Necula, G., Paszke, A., Vander{P}las, J., Wanderman-{M}ilne, S., and Zhang, Q., ``{JAX}: composable transformations of {P}ython+{N}um{P}y programs,'' (2018).

\bibitem{jaxopt_implicit_diff}
Blondel, M., Berthet, Q., Cuturi, M., Frostig, R., Hoyer, S., Llinares-L{\'o}pez, F., Pedregosa, F., and Vert, J.-P., ``Efficient and modular implicit differentiation,'' {\em arXiv preprint arXiv:2105.15183}  (2021).

\bibitem{douglas2023}
Douglas, E.~S., Martin, B., Males, J.~R., Aldering, G., Kim, D., Perlmutter, S., and Jannuzi, B., ``Taking big risks for big science: approaches to lowering the cost of large space telescope,'' in [{\em Astronomical Optics: Design, Manufacture, and Test of Space and Ground Systems IV}{\nolinebreak\hspace{0.1em}]},  SPIE (2023).

\end{thebibliography}
\bibliographystyle{spiebib} % makes bibtex use spiebib.bst

\end{document}